\documentclass[superscriptaddress, prl, twocolumn]{revtex4}

\usepackage{graphicx}
\usepackage{verbatim}
\usepackage{amsmath,amssymb,relsize}

\makeatletter
\def\Ddots{\mathinner{\mkern1mu\raise\p@
\vbox{\kern7\p@\hbox{.}}\mkern2mu
\raise4\p@\hbox{.}\mkern2mu\raise7\p@\hbox{.}\mkern1mu}}
\makeatother
\usepackage{amsmath}
\usepackage{amsthm}
\usepackage{amssymb}
\usepackage{rotating}
\usepackage{epstopdf}
\usepackage{bbold}
\usepackage{times}
\usepackage[none]{hyphenat}
\usepackage{color}

\begin{document}

\author{G. Pica}
\affiliation{SUPA, School of Physics and Astronomy, University of St Andrews, KY16 9SS, United Kingdom}
\author{B. W. Lovett}
\affiliation{SUPA, School of Physics and Astronomy, University of St Andrews, KY16 9SS, United Kingdom}
\author{R. N. Bhatt}
\affiliation{Dept. of Electrical Engineering, Princeton University, Princeton, New Jersey 08544, USA}
\author{S. A. Lyon}
\affiliation{Dept. of Electrical Engineering, Princeton University, Princeton, New Jersey 08544, USA}

\title{Exchange coupling between silicon donors: the crucial role of the central cell and mass anisotropy}

\begin{abstract}
Donors in silicon are now demonstrated as one of the leading candidates for implementing qubits and quantum information processing. Single qubit operations, measurements and long coherence times are firmly established, but progress on controlling two qubit interactions has been slower. One reason for this is that the inter donor exchange coupling has been predicted to oscillate with separation, making it hard to estimate in device designs. We present a multivalley effective mass theory of a donor pair in silicon, including both a central cell potential and the effective mass anisotropy intrinsic in the Si conduction band. We are able to accurately describe the single donor properties of valley-orbit coupling and the spatial extent of donor wave functions, highlighting the importance of fitting measured values of hyperfine coupling and the orbital energy of the $1s$ levels. Ours is a simple framework that can be applied flexibly to a range of experimental scenarios, but it is nonetheless able to provide fast and reliable predictions. We use it to estimate the exchange coupling between two donor electrons and we find a smoothing of its expected oscillations, and predict a monotonic dependence on separation if two donors are spaced precisely along the [100] direction. 
\end{abstract}

\maketitle

\section{I. Introduction} It is now fifteen years since Kane proposed his blueprint for building a quantum computer using phosphorus donors in silicon (Si:P)~\cite{kane98}. After years of steady progress towards realising this dream, recent remarkable experiments on uncoupled donors have brought it much closer to reality. The longest nuclear spin coherence time for (Si:P) now exceeds an astonishing thirty-nine minutes at room temperature~\cite{saeedi13}, and electron spin coherence survives for more than one second~\cite{tyryshkin12}. It has also been possible to measure~\cite{morello10} and manipulate~\cite{pla13} an individual P-donor nuclear spin. However, still lacking is a way of controllably coupling multiple donors together to generate the kinds of correlated quantum states required for universal quantum information processing.

\begin{figure}[h!]
  \centering
    \includegraphics[width=.5\textwidth]{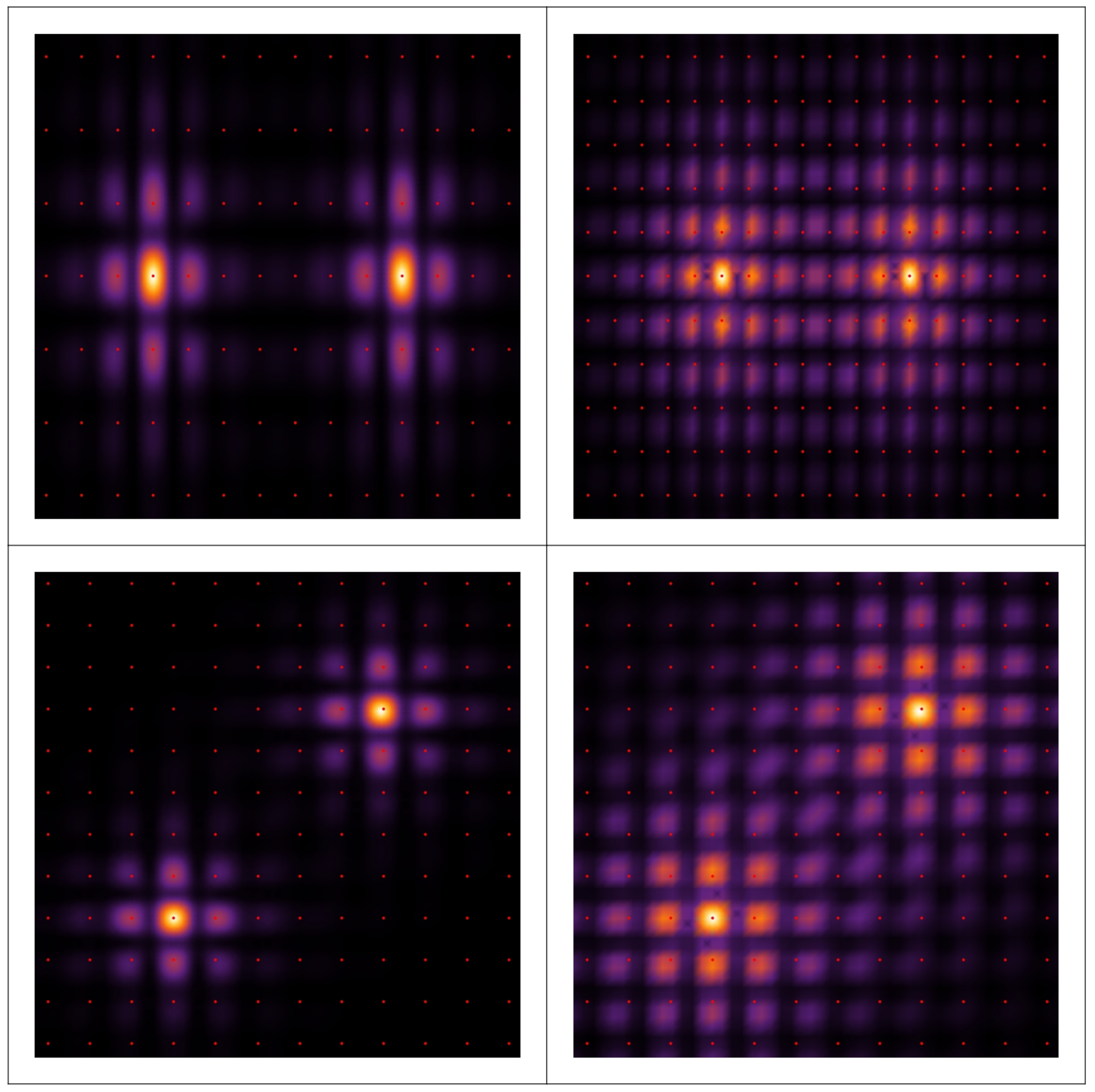}
  \caption{Plots of the spatial electronic densities around two adjacent implanted donor nuclei, in a plane containing the vector separation $\textbf{d}$. The two panels above refer to $\textbf{d}$ along [100], those below to [110]. Left panels are calculated with the wave functions used in the MV EMT theory, and show stronger concentration of the density around nuclei (hence larger hyperfine coupling) than the right panels which use KL wave functions \cite{kohnlut}. Red dots highlight the positions of the Si nuclei of the underlying lattice. The mismatch between their locations and the local critical points of the density is a result of the nontrivial structure of the Si conduction band.}
  \label{densityplots}
\end{figure}

Perhaps the most conceptually straightforward way of coupling two donors together is exactly as Kane proposed: to place two donors closely enough that their electronic wave functions overlap (Fig. \ref{densityplots}). This results in an interaction between donors that is Coulombic in nature, and depends strongly on the electronic density of both donors involved. The spatial region which gives the largest contribution to the interaction is concentrated around the inter-donor separation axis, midway between the nuclei; varying the potential of a surface electrostatic gate may then modulate this overlap \cite{externalexchange}, enabling a controllable switching of the donors' coupling. A critical question then is how large the coupling can be, and how accurately the donors must be placed for gates to be robust to variations in the coupling strength. 

Previous work employing an effective mass theory of Si:P \cite{oscilla,oscilla2,zalba} predicts strong oscillations in the dependence of the coupling on distance, and is based on earlier work \cite{bhatt} which investigated the effect of valley-mixing in multi-valley semiconductors like silicon. This may cause larger changes in coupling strength in silicon, as a donor is moved from one lattice site to the next, than would be expected for semiconductors with non-degenerate conduction band minima. 

Such a situation represents a tough experimental challenge since it suggests that donors need to be placed very accurately. A more recent numerical calculation \cite{w&h} going beyond effective mass theory finds that the oscillations are suppressed relative to the Kohn-Luttinger effective mass approximation (KL). In this work, we consider the donor problem within a multi-valley effective mass theory (MV EMT) including effects of both the central cell and mass anisotropy present for Si:P. Our approach not only allows the entire $1s$ manifold of the binding energies of the isolated donor electron to be accurately described, but crucially it also allows a correct description of the hyperfine coupling to the donor nucleus, as measured in experiments. The most important consequence is our finding that the spread of the electronic wavefunction was significantly overestimated by previous treatments, which only relied on fitting of orbital energies. Building on this result, we will show how the anisotropy of the donor wave function leads to a suppression in the oscillatory nature of the exchange coupling, especially for certain geometries. 
Similar effects have been previously predicted~\cite{w&h},  but we
are able to clarify their origin and to improve their evaluation through correct fitting of the short-range behaviour of the donor wave function. This improvement in turn modifies intermediate and long range wave function shape and extent, and so strongly influences our exchange coupling estimates.
In addition, our theory is much less numerically intensive and easily adaptable to more complicated electromagnetic environments and different donor species.

In the next section, we survey previous theoretical approaches of the same problem, and derive the multi-valley EMT equation we are going to use. We discuss known limitations of our theory and explain why it is still useful for our purposes, and then we fit experimental quantities of interest with only two free parameters. In section III, we discuss the method used for direct evaluation of donor-donor exchange coupling as a function of donor separation: and present our results. We are able to provide a clear physical explanation for our predictions, based on the analytic nature of our treatment, and point out interesting consequences for experimental implementations. We conclude in Section IV.

\section{II. Theory} We start with the Hamiltonian:
\begin{equation}\label{eqbase}
H \Psi(\textbf{r})= [-\frac{\hbar^{2}}{2 m_{0}}\nabla^{2} + V^{0}(\textbf{r}) + U(\textbf{r}) ]\Psi(\textbf{r})= E \Psi(\textbf{r}),
\end{equation}   
where $\Psi(\textbf{r})$ is the wave function of the donor electron, $m_{0}$ is its rest mass, $V^{0}(\textbf{r})$ is the periodic potential of the undoped silicon crystal, and $U(\textbf{r})$ accounts for the interaction with the impurity ion. $E$ stands for the resulting energy eigenvalues.

The exact solution to the undoped Si case, with $U(\textbf{r})=0$, is provided by the Bloch functions $\phi_{n}(\textbf{k},\textbf{r})=u_{n}(\textbf{k},\textbf{r})e^{i \textbf{k} \cdot \textbf{r}}$ \cite{yucardona} (where $n$ is the band index, $\textbf{k}$ lies within the first Brillouin zone, and $u_{n}(\textbf{k},\textbf{r})$ has the periodicity of the lattice). Then, bound donor electrons can be described with $U(\textbf{r})=- \frac{e^{2}}{\epsilon_{Si}|\textbf{r}|}$, i.e. a screened attractive Coulomb interaction with the extra positive charge of the substitutional impurity. Since the effective Bohr radius of shallow donor bound states is expected to be considerably larger than the lattice spacing, Kohn and Luttinger \cite{originalemt} showed that it is reasonable to write $\Psi(\textbf{r})$ as an expansion in the Bloch states above, restricted to the deepest conduction band ($n=0$). Included in their wave function were only the Bloch states around one of the six degenerate minima in silicon;  these `valleys' are placed at $\textbf{k}_{0\mu}=\frac{2 \pi }{a_{Si}}0.85(\pm \hat{\textbf{x}},\pm \hat{\textbf{y}},\pm \hat{\textbf{z}})$ ($a_{Si}=5.43$\AA~is the silicon lattice constant). The coefficients of such expansion are called EMT envelopes, and satisfy a simpler equation than the full Eq.~\ref{eqbase}. Even though the new equation is not analytically solvable, an excellent approximation to the exact solution can be achieved via variational minimization of the expectation value of the energy. Such approximations, though, fail to describe the $s$-donor levels, and especially the ground electronic state, because the short-range impurity potential unique to each chemical donor species - the so called `central cell' correction~\cite{ningsah} - is not properly captured. Moreover, this potential shows strong variations within the unit cell surrounding the impurity nucleus, so that all the six valleys are coupled (valley-orbit interactions). 
For this reason we use a multi-valley EMT and expand $\Psi(\textbf{r})$ in terms of the Bloch functions close (in $k$-space) to each of those six minima \cite{hui}:
\begin{equation}\label{expansion}
\Psi(\textbf{r})=\sum_{\mu=1}^{6}\alpha_{\mu}\frac{1}{(2 \pi)^{3}} \int \tilde{F}_{\mu}(\textbf{k}_{\mu}+\textbf{k}_{0\mu})\phi_{0}(\textbf{k}_{\mu}+\textbf{k}_{0\mu},\textbf{r}) d\textbf{k}_{\mu},
\end{equation}  
where $\tilde{F}_{\mu}(\textbf{k}_{\mu}+\textbf{k}_{0\mu})$ is the expansion coefficient for the Bloch function $\phi_{0}(\textbf{k}_{\mu}+\textbf{k}_{0\mu},\textbf{r})$ centered around the $\mu$th valley.

Even without precise knowledge of $U(\textbf{r})$, the residual $T_{d}$ symmetry of the system dictates the shape of the eigenstates into which the previously degenerate $1s$ ground state is split: a singlet $A_{1}$, a triplet $T_{2}$, and a doublet $E$. The singlet is an equal symmetric superposition of all six valleys ($\{\alpha_{i}(A_{1})\}=\frac{1}{\sqrt{6}}(1,1,1,1,1,1)$), with the other states forming orthogonal combinations of the $\{\alpha_{i}\}$.

We now take the expectation value of Eq.~\ref{eqbase} with respect to Eq.~\ref{expansion} \cite{shindonara} and go through the usual EMT approximations \cite{hui,w&h} to obtain
\begin{eqnarray}
\begin{aligned}\label{intermezzo}
0=\int d\textbf{r} \sum_{\mu=1}^{6}\alpha^{\ast}_{\mu}F^{\ast}_{\mu}(\textbf{r}) [\alpha_{\mu} (\textbf{p}\cdot \textbf{A}_{\mu}\cdot \textbf{p}-E)F_{\mu}(\textbf{r}) + \\
\sum_{\nu=1}^{6} \alpha_{\nu} e^{-i(\textbf{k}_{0\mu}-\textbf{k}_{0\nu})\cdot \textbf{r}} u_{0}^{\ast}(\textbf{k}_{0\mu},\textbf{r})u_{0}(\textbf{k}_{0\nu},\textbf{r}) U(\textbf{r}) F_{\nu}(\textbf{r}) ] ,
\end{aligned}
\end{eqnarray}
where $\textbf{A}_{\mu}$ is the anisotropic inverse effective mass tensor for silicon, which describes the curvature of bands parallel and perpendicular to the wave vector locating each of the band minima within the Brillouin Zone: $m^{\ast}_{\perp}=0.191 m_{0}$ and $m^{\ast}_{\parallel}=0.916 m_{0}$. Using $u_{0}^{\ast}(\textbf{k},\textbf{r})u_{0}(\textbf{k}',\textbf{r})=\sum_{\textbf{G}}C_{\textbf{G}}(\textbf{k},\textbf{k}') e^{i \textbf{G}\cdot \textbf{r}}$ (where $\textbf{G}$ runs over the vectors of the silicon reciprocal lattice) \cite{originalemt}, and neglecting the  $\textbf{G}\neq 0$ terms \footnote{
This approximation has been criticized for example in \cite{resta} and \cite{w&h} with the argument that there are some Fourier components of the impurity potential $U(\textbf{q})$ on the scale of $\textbf{q}=\textbf{k}-\textbf{k}'-\textbf{G}$ with $\textbf{G}\neq 0$ which are actually more important than those with $\textbf{G}= 0$ with the same $\textbf{k}$ and $\textbf{k}'$. However, some counter-arguments can be provided: $i)$ as pointed out in \cite{resta}, neglecting the Umklapp components of the pseudopotential can only result in an underestimation of some inter-valley matrix elements, i.e. $\textbf{k}\neq \textbf{k}'$, which are considerably smaller than the intra-valley ones ($\textbf{k}= \textbf{k}'$); $ii)$ the statistical weight of those wave vectors over which the envelopes $F_{\mu}$ are significant is small, and the relative matrix element comes from an integration over the whole of $k_{\mu}$ space; $iii$) the neglected contributions are further depressed by $C_{\textbf{G}}(\textbf{k}_{0q},\textbf{k}_{0p})$ with $\textbf{G} \neq 0$ which, although not exactly known, should be smaller than the respective quantities with $\textbf{G}=0$ as derived in using some theoretical treatments (see for example \cite{rescaresta}).
} leads to:
\begin{eqnarray}\label{definitiva}
\begin{aligned}
\int d\textbf{r} \sum_{\mu=1}^{6}\alpha^{\ast}_{\mu}F^{\ast}_{\mu}(\textbf{r}) \times [\alpha_{\mu} (\textbf{p}\cdot \textbf{A}_{\mu}\cdot \textbf{p}-E)F_{\mu}(\textbf{r}) +\\
\sum_{\nu=1}^{6} \alpha_{\nu} e^{-i(\textbf{k}_{0\mu}-\textbf{k}_{0\nu})\cdot \textbf{r}} C_{\textbf{0}}(\textbf{k}_{0\mu},\textbf{k}_{0\nu}) U(\textbf{r}) F_{\nu}(\textbf{r})] =0  
\end{aligned}
\end{eqnarray}
where 
$C_{0}(\textbf{k}_{0q},\textbf{k}_{0q}) = 1,  C_{0}(\textbf{k}_{0q},-\textbf{k}_{0q})= -0.1728 $ and $ C_{0}(\textbf{k}_{0q},\textbf{k}_{0\pm p}) = 0.4081$ ($p \neq q$), values
taken from a calculation with the pseudo-potential form factors of the periodic undoped silicon crystal \cite{cohen}, performed by Shindo and Nara \cite{shindonara}
to describe its band structure.

We use the impurity potential first proposed by Ning and Sah \cite{ningsah}: 
\begin{equation}
U(\textbf{r})=-\frac{e^{2}}{\epsilon_{Si}|\textbf{r}|}(1-\text{e} ^{- b |\textbf{r}|}+ B |\textbf{r}| \text{e} ^{- b |\textbf{r}|}).
\end{equation}
$b$ and $B$ are parameters that are fit to experimental data. The potential resembles the screened hydrogenic Coulomb interaction at large distances, while at extremely short range it mimics the extra nuclear charges embedded in the substitutional donor impurity. In essence, the potential is an average of the oscillations over the central-cell lengthscale with a phenomenological model potential $U(\textbf{r})$ that still satisfies the EMT assumptions above 
- most importantly, smoothness \cite{shindonara} -
 but gives a good description of the experimentally determined valley-orbit energies. Our method proceeds as follows: for each trial calculation, we first fix $b$ and $B$ and then we use the following ansatz for the envelopes \cite{kohnlut}, e.g. 
\begin{eqnarray}\label{trials}\nonumber
F_{\pm z}= \sqrt{\frac{1}{\pi a_{D}^{2}b_{D}}} \exp\left(-\sqrt{\frac{x^{2}+y^{2}}{a^{2}_{D}}+\frac{z^{2}}{b^{2}_{D}}}\right),
\\
F_{\pm x}= \sqrt{\frac{1}{\pi a_{D}^{2}b_{D}}} \exp\left(-\sqrt{\frac{z^{2}+y^{2}}{a^{2}_{D}}+\frac{x^{2}}{b^{2}_{D}}}\right).
\end{eqnarray} 
We now minimize the expectation values of the energies of the three split $1s$ levels according to equation (\ref{definitiva}) by varying $a_{D}$ and $b_{D}$ separately for each. We then find the best values of $b$ and $B$ by finding a good match between our predictions and measured ground state donor energy~\cite{aggarwal} and hyperfine coupling \cite{feher,fletcher} for Si:P.
We emphasize that, unlike some previous multi-valley EMT treatments, the envelopes we have used have the crucial property of anisotropy. This is vital for calculations of properties of a donor-donor system, which clearly has a broken symmetry along the vector connecting the two donors. Isotropic envelopes provide predictions of exchange coupling that can be qualitatively different to those we present here.

With $b=19.96$nm$^{-1}$ and $B=246.1$nm$^{-1}$ we obtain $E_{A_{1}}=-45.5$meV, $E_{T_{2}}=-36.0$meV, $E_{E}=-33.0$meV, which must be compared with the experimental \cite{aggarwal} $E_{A_{1}}=-45.57$meV, $E_{T_{2}}=-33.74$meV, and $E_{E}=-32.37$meV: other than the fitted singlet, the agreement is very good for the doublet, and somewhat less accurate for the triplet, but not unacceptably so. 
In addition, we can fit the value of the squared electron wave function at the donor nucleus  $|\psi(0)|^{2}$, which is proportional to the hyperfine coupling between the impurity nucleus and the donor electron, by expressing it as $|\psi(0)|^{2}\approx 6 \eta |F(0)|^{2}$. Here $\eta=|u_{0}(\textbf{k}_{0},0)|^{2}/\langle |u_{0}(\textbf{k}_{0},\textbf{r})|^{2}\rangle_{\text{unit cell}} =186\pm 18$. We set this to match the $|\Psi_{A_{1}}(0)|^{2}=4.4\times 10^{29}$m$^{-3}$ \cite{fletcher} extracted from experimental measurements of the hyperfine constant.
\footnote{Different $\eta$ values, proposed by other experiments and theory \cite{assali}, would lead to slightly different Bohr radii (about 5\%), what would not change any of the conclusions presented later.}
    
\section{III. Donor-donor exchange}
We used the Heitler-London (HL) approach \cite{hl} to evaluate the exchange splitting between two adjacent P donor electrons in a Si layer. HL uses a smart guess of the ground and first excited molecular orbital states of the two-particle system, based on single particle ground state orbitals. The two resulting states have a difference in energy of $J=E_{T}-E_{S}$, where the spin-singlet $|S\rangle = \frac{1}{\sqrt{2}}|\hspace{-1.5mm}\uparrow \downarrow -\hspace{-1.5mm} \downarrow \uparrow\rangle$ and the spin-triplet $T_{0,+,-}=\frac{1}{\sqrt{2}}|\hspace{-1.3mm} \uparrow \downarrow +\hspace{-1.3mm}  \downarrow \uparrow\rangle, |\hspace{-1mm} \uparrow\uparrow\rangle, |\hspace{-1.5mm} \downarrow\downarrow\rangle $ have spatial wave functions made up of a symmetric and an antisymmetric combination of the single particle orbitals respectively. The convention we have chosen for the sign of the $H$ ensures that $J$ is positive; this must indeed be the case at zero magnetic field, by the Lieb-Mattis theorem \cite{mattis}. This Coulomb interaction can then be effectively described as a spin Hamiltonian term coupling the two donor electrons:
\begin{equation}
H=J \textbf{S}_{1}\cdot\textbf{S}_{2}.
\end{equation}
The evaluation of exchange coupling has been attempted by several different theoretical approaches.
Andres {\it et al.} \cite{bhatt} first emphasized that, unlike the monotonic decay of $J(d)$ characteristic of the H$_{2}$ molecule in vacuum, the exchange coupling was expected to show oscillations as a function of the donor separation, because the conduction-band minima in Si are away from the Brillouin zone center, and the corresponding Bloch functions interfere with one another. As further pointed out in \cite{oscilla} and \cite{oscilla2}, this can lead to serious difficulties when trying to harness exchange coupling for quantum computation. The resolution of the donor positioning during the implantation process is not refined enough to ensure that all donors within a Si layer would experience even the same order of magnitude of $J$. 

In \cite{w&h} a numerical solution of the full Hamiltonian describing the donor electron is proposed, obtained through exact diagonalization in the basis of the undoped crystal Bloch functions - the Band Minima Basis method (BMB). Predictions made using this method are not limited by any of the EMT approximations, but only by the convergence and numerical accuracy of the computation, and by the validity of the pseudopotential used \cite{pantelides}. 
Such detailed and numerically
intensive microscopic calculations predict that the strength of the coupling is reduced, and its oscillations have their amplitude decreased as compared to the calculations performed with the multi-valley wave function involving Kohn and Luttinger envelopes \cite{originalemt}. This happens since the KL approach includes the correct valley structure without taking into account its consequences on the donor Hamiltonian, i.e. central cell corrections. However, all calculations so far still fail to get a reliable description of the electronic density in the region close to the donor nucleus.

\begin{figure}
  \centering
    \includegraphics[width=.5\textwidth]{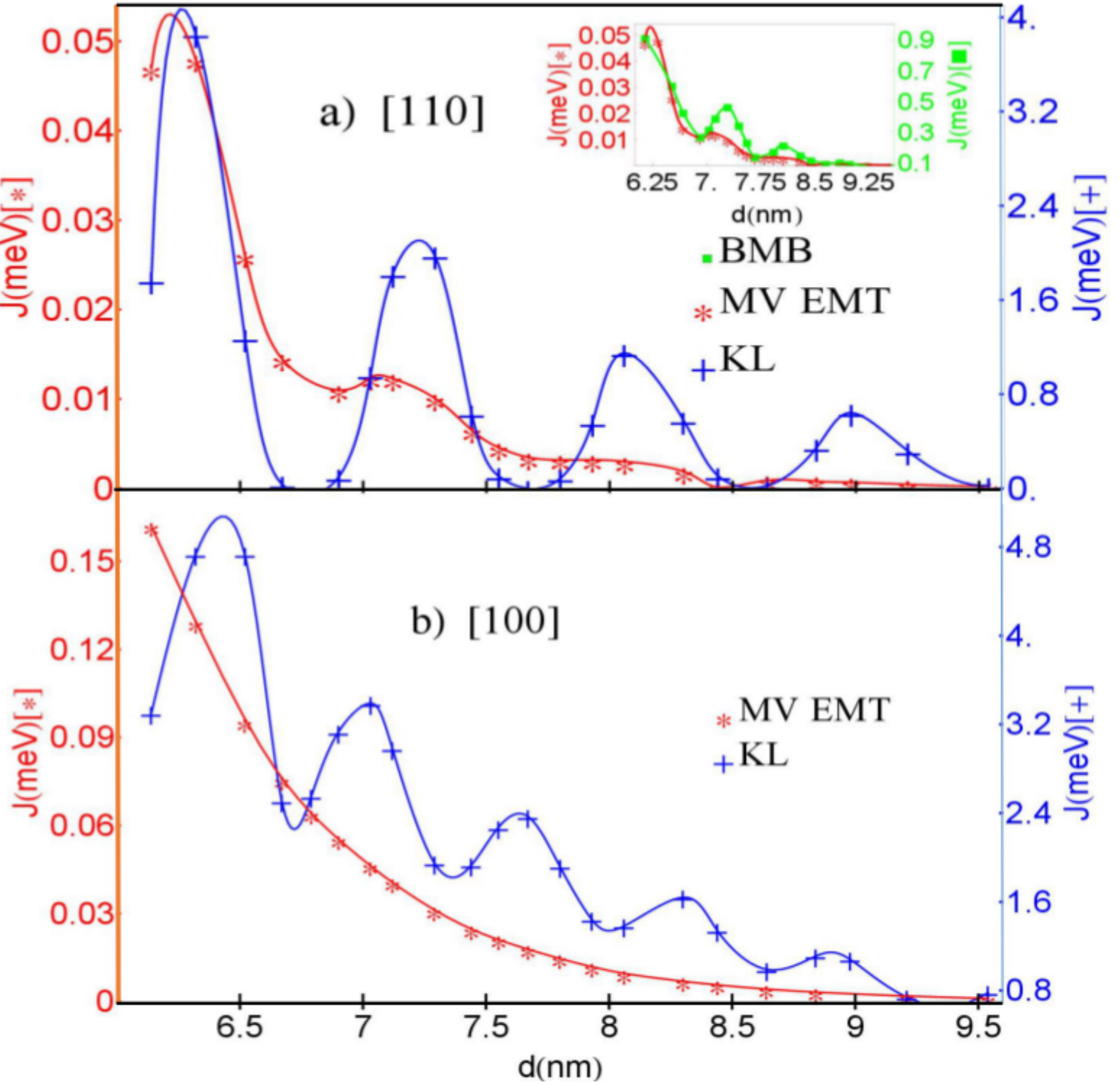}
  \caption{The exchange splitting between electrons pertaining to adjacent Si:P donors is shown as a function of their separation $d$ along (a) the [110] axis and (b) the [100] axis $b)$. The range of $d$ displayed refers to the realistic uncertainty in the resolution of the placement of donors in the Si layer during the implantation process. The solid lines are only a guide to the eye, providing a rough interpolation between the calculated data points. We display both the KL solution of the donor Hamiltonian (blue crosses) and our MV EMT (red stars). The large difference in the scale of the two y-axes makes apparent the discrepancy between the two calculations. The inset in (a) shows the juxtaposition of our results with those obtained from BMB (green squares), extracted from \cite{w&h}. The most striking property of the $d$ dependence we have calculated for the [100] axis is the absence of the oscillations expected from other theories~\cite{oscilla,oscilla2}, a finding explained in detail in the text.}
  \label{wash}
\end{figure}

We now explore two donor coupling with our new MV EMT which, unlike the KL method, does account for the effect of central cell corrections on the donor, and does accurately predict single particle properties.
The two-particle integrals entering $J$ were computed with a fast Monte-Carlo algorithm for adaptive multidimensional integration (cubature) \cite{johnson}, and each data point takes only a few minutes to compute. Fig.~\ref{wash} shows our evaluation of $J$ for donor separation $d$ in the [110] and [100] spectroscopic directions, compared with corresponding values we determined using KL \cite{kohnlut}.
The biggest difference between the two theories lies in the magnitude of the exchange splittings: the extra localization in real space provided by the strong short range potential of the impurity for MV EMT leads to a shrinking of the effective Bohr radii of the ground state envelopes ($a_{D}=1.15$nm, $ b_{D}=0.61$nm), when compared to KL ($a_{D}=2.509$nm, $ b_{D}=1.443$nm \cite{oscilla2}). This is illustrated by the electron density plots shown in Fig.~\ref{densityplots}.

For the [110] direction, the same qualitative behavior predicted in \cite{w&h} is apparent, but we find shallower oscillations. To explain this, consider the following approximation of the whole exchange splitting calculated here, the so-called indirect exchange integral \cite{bhatt}. It has the advantage of clear analytical structure, and the has the same qualitative behaviour as $J(\textbf{d})$:
\begin{equation}
j(\textbf{d})=\sum_{\mu,\nu}|\alpha_{\mu}|^{2}|\alpha_{\nu}|^{2}j_{\mu\nu}(\textbf{d})\cos(\textbf{k}_{0\mu}-\textbf{k}_{0\nu})\cdot \textbf{d}.
\label{eq:exchange}
\end{equation}
where $j_{\mu\nu}(\textbf{d})$ is the indirect exchange integral between the envelopes $F_{\mu}(\textbf{r}_{1})$ and $F_{\nu}(\textbf{r}_{2})$:
\begin{equation}\label{simpler}
\int d\textbf{r}_{1}d\textbf{r}_{2}F^{\ast}_{\mu}(\textbf{r}_{1})F^{\ast}_{\nu}(\textbf{r}_{2}-\textbf{d})\frac{e^{2}}{\epsilon |\textbf{r}_{1}-\textbf{r}_{2}|}F_{\mu}(\textbf{r}_{1}-\textbf{d})F_{\nu}(\textbf{r}_{2})
\end{equation}

The sinusoidal terms arise from the periodic parts of the Bloch states. The \emph{longitudinal} $j_{\mu \nu}^{l}$, where either $\textbf{k}_{0\mu}$ or $\textbf{k}_{0\nu}$ has some component along $\textbf{d}$, give oscillating contributions to $j(\textbf{d})$ and are responsible for the large oscillations apparent in the KL (and BMB) cases. The \emph{transverse} $j_{\mu\nu}^{t}$ where $\textbf {k}_{0 \mu}\cdot \textbf{d}=\textbf {k}_{0 \nu}\cdot \textbf{d}=0$ decrease monotonically with $d$.

Owing to the large difference between longitudinal and transverse effective masses in Si used in MV EMT, our envelopes are very anisotropic: we get $a_{D}/b_{D}\approx 1.90$, compared to KL's 1.74. To explain why anisotropy gives a great suppression of the oscillating terms in MV EMT, we introduce the
two-envelope overlap integral ${\cal S}(\textbf{d})=\int d\textbf{r} \Psi(\textbf{r}) \Psi(\textbf{r}-\textbf{d})$. Both the envelope overlap parts of the $\bf{r}_1$ and $\bf{r}_2$ integrands in Eq.~\ref{simpler} are peaked in the region between the two donors - i.e. for the same values of $\bf{r}_1$ and $\bf{r}_2$. The denominator of the integrand has its largest value when $\bf{r}_1-\bf{r}_2$ is small; it can therefore be shown that the $d$ dependence of the exchange integral,~Eq.~\ref{eq:exchange}, is dominated by that of ${\cal S}^2(\bf d)$\cite{oscilla}. This is true so long as $\{a_{D},b_{D}\}/d$ are small enough, which they are for all results presented here.

It can be shown \cite{oscilla} that 
\begin{equation}\label{overlap}
{\cal S}(\textbf{d})\approx \sum_{\mu}{\cal S}_{\mu\mu}(\textbf{d}) \equiv \sum_{\mu}|\alpha_{\mu}|^{2} \text{e}^{-i \textbf{k}_{\mu}\cdot \textbf{d}} \text{e}^{- \mathit{d}_{\mu}}(1+\mathit{d}_{\mu} + \mathit{d}_{\mu}^{2}/3)
\end{equation}
where $\textbf{\textit{d}}_{\mu}$ is the separation vector $\textbf{d}$ appropriately rescaled with the anisotropic Bohr radii: $b_{D}$ along $\hat{\mu}$ direction, $a_{D}$ for the others, e.g. $\textbf{\textit{d}}_{z}=(d_{x}/a_{D},d_{y}/a_{D},d_{z}/b_{D})$. For the range of separations explored, the decaying exponential term dominates the functional dependence of the ${\cal S}$. For example, with $\textbf{d}$ parallel to [110]
\begin{equation}
|\frac{\partial}{\partial d} \log({\cal S}^{l}_{\mu\mu})|/|\frac{\partial}{\partial d} \log({\cal S}^{t}_{\nu\nu})|\propto \frac{\sqrt{a_D^2+b_D^2}}{b_D\sqrt{2}}>1 .
\end{equation}  
Hence the oscillating longitudinal terms decay more quickly with $d$ than the transverse ones; as $d$ increases, oscillations are smoothed out. Anisotropy plays a  key role in this effect, and this is far more evident within MV EMT, where our fitting of hyperfine coupling results in a spread of the donor wave functions that is much smaller than those in KL or BMB.
With $\textbf{d}$ directed along the [110] direction, even though 32 of the 36 terms in Eq.~\ref{eq:exchange} are transverse, these are heavily suppressed and the oscillations then appear shallower in MV EMT than in the other theories.

Even more striking is the form of $J(\textbf{d})$ when the separation lies precisely along the [100] direction (see panel $b)$ of figure \ref{wash}). Thanks to the higher symmetry in this case, only four of the 36 $j_{\mu\nu}(\textbf{d})$ are associated with oscillations, and these are suppressed to such an extent that the exchange is now monotonically decreasing.

\section{IV. Conclusions}
We have presented a theoretical analysis of the P-donor electron wave function in Si. Our consistency with the measured hyperfine interaction strength improves the description of the electronic density in the region between neighbouring donor nuclei, which determines their exchange coupling.
Ours is a relatively simple and numerically light framework, which nonetheless is able to reliably predict properties of shallow electronic states in silicon.
The limitations and approximations of our theory are clearly understood, and possible improvements may come from an exact knowledge of the Si Bloch eigenfunctions and the short-range potential characteristic of each donor: both still are inaccessible even with \emph{ab initio} approaches.
We have shown why the anisotropy intrinsic to the Si conduction band is particularly important for estimating the exchange splitting within the cylindrically symmetric two-donor system: the most immediate consequence is the large difference in the distance dependence of $J$ for donors separated along different spectroscopic axes. We find the same qualitative effect of `washing out' of the oscillations in $J(d)$ as in the \emph{ab initio} calculations in \cite{w&h}, but the size of the exchange and the amplitude of oscillations are significantly reduced.
Precisely along the [100] direction, we predict that there will be no oscillations at all in the dependence of $J$ on separation. The reasoning outlined at the end of the previous section allows us to anticipate that oscillations will be smoothed efficiently at smaller distances the more localized the impurity electron. Thus at fixed donor-separation, the predicted effect will be more pronounced for As, Sb and Bi-implanted silicon. Even though oscillatory variations of $J$ are still expected as a function of misplacements orthogonal to a nominal [100] separation (those trends would resemble qualitatively the $J$ dependence on inter-donor separations along [110] and [111] directions), the range of interaction strengths induced by uncertainty in donor implantation position will be less than previously thought. 
Future work will explore extensions of MV EMT to include the effects of external electric or magnetic fields, and other dopants.

\section{Acknowledgements}
This research was funded by the joint EPSRC (EP/I035536) / NSF (DMR-1107606) Materials World Network grant (BWL, GP, SAL), partly by the NSF MRSEC grant DMR-0819860 (SAL), the Department of Energy, Office of Basic Energy Sciences grant DE-SC0002140 (RNB). BWL thanks the Royal Society for a University Research Fellowship and RNB thanks the Institute for Advanced Study, Princeton for hospitality during the period this work was written.

\end{document}